\documentclass[aps,preprint,showpacs,preprintnumbers,amsmath,amssymb,floatfix]{revtex4}
\usepackage{graphicx}
\begin{document}
\title{Exact quantum dynamics of a bosonic Josephson junction}
\author{Kaspar Sakmann, Alexej I. Streltsov, Ofir E. Alon and Lorenz S. Cederbaum}
\affiliation{Theoretische Chemie, Physikalisch-Chemisches Institut, Universit\"at Heidelberg,\\
Im Neuenheimer Feld 229, D-69120 Heidelberg, Germany}
\begin{abstract}
The quantum dynamics of a one-dimensional bosonic Josephson junction 
is studied by solving the time-dependent 
many-boson Schr\"odinger equation numerically exactly.
Already for weak interparticle interactions and on short 
time scales, the commonly-employed mean-field 
and many-body methods are found to deviate 
substantially from the exact dynamics.
The system exhibits rich many-body dynamics like enhanced 
tunneling and a novel equilibration phenomenon 
of the junction depending on the interaction, 
attributed to a quick loss of coherence.
\end{abstract}

\maketitle
Recent experiments on interacting 
Bose-Einstein condensates 
in double-well traps have led to some of 
the most exciting results
in quantum physics, including matter-wave 
interferometry \cite{e4,e8}, squeezing and entanglement \cite{e35,e9} 
as well as work on high-precision sensors \cite{e10}.
Particular attention has been paid to tunneling phenomena of 
interacting Bose-Einstein condensates in double-wells, 
which in this context are referred to as bosonic Josephson junctions.
Explicitly, Josephson oscillations and self-trapping (suppression 
of tunneling) with Bose-Einstein condensates 
have been predicted \cite{t1,t2} and recently 
realized in experiments \cite{e1,e2},  
drawing intensive interest, see, e.g., \cite{2mode1,clee,
2mode2,rgat,meso,2mode3} and references therein.

For the first time in literature  we provide 
the numerically exact many-body
quantum dynamics of a one-dimensional (1D) bosonic 
Josephson junction in this work.
This is made possible by a breakthrough 
in the solution of the time-dependent 
many-boson Schr\"odinger equation.
We use the exact solution to check the current
understanding of bosonic Josephson junctions -- 
commonly described by the popular 
Gross-Pitaevskii (GP) mean-field theory
and the Bose-Hubbard (BH) many-body model -- 
and to find novel phenomena.
The results of the GP and BH theories
are found 
to deviate substantially from the full many-body solution,
already for weak interactions and on short time scales.
In particular, the well-known self-trapping effect 
is greatly reduced.
We attribute these findings to a quick loss of the junction's coherence 
not captured by the common methods.
For stronger interactions and on longer time scales, 
we find a novel equilibration dynamics in which the
density and other observables of the junction tend 
towards stationary values. We show that the dynamics of 
bosonic Josephson junctions is much richer 
than what is currently known.

To compute the time evolution of the system,
we solve the time-dependent many-boson Schr\"odinger equation by
using the multiconfigurational time-dependent 
Hartree for bosons (MCTDHB) method \cite{MCTDHB}.
In the MCTDHB($M$) method the time-dependent many-boson wavefunction
is expanded in all time-dependent permanents $|\vec{n};t\rangle$,
generated by distributing $N$ bosons over $M$ 
time-dependent orbitals $\{\phi_i(x,t)\}$.
$\vec{n}=(n_1,n_2,\cdots n_M)$ collects the occupation numbers.
The MCTDHB wavefunction thus reads
$|\Psi(t)\rangle = \sum_{\vec{n}} C_{\vec{n}}(t) |\vec{n};t\rangle$.
The expansion coefficients $\{C_{\vec{n}}(t)\}$ 
{\em and} the orbitals $\{\phi_i(x,t)\}$ 
are determined by the Dirac-Frenkel
time-dependent variational principle \cite{MCTDHB}.
The present results are obtained by using a novel 
mapping of the many-boson configuration space in 
combination with a parallel implementation of MCTDHB,
allowing the efficient handling of millions of time-dependent,
optimized permanents \cite{parll}.
We note that GP theory is contained
in the MCTDHB framework as the special case  $M=1$. 
The many-body wavefunction then becomes a single permanent
and the dynamics is restricted to remaining condensed at all times.
The BH model for this system employs 
two orbitals and is thus, in principle, 
capable of describing correlations. However, as we shall see below,
the fact that the BH orbitals are {\it time-independent} causes 
the BH model to underestimate correlations. 

With the time-dependent many-boson wavefunction 
$|\Psi(t)\rangle$ at hand, any quantity of interest 
of the interacting many-boson system can be computed.
Here we focus on the evolution of the following quantities 
to analyze the dynamics of the Josephson junction.
The reduced one-body density matrix of the system
is defined by 
$\rho^{(1)}(x|x';t)=\langle\Psi(t)|\hat{\mathbf \Psi}^\dag(x')\hat{\mathbf \Psi}(x)|\Psi(t)\rangle$,
where $\hat{\mathbf \Psi}(x)$ is the usual bosonic 
field operator annihilating a particle at position $x$.
Its diagonal part, $\rho(x,t)\equiv\rho^{(1)}(x|x'=x;t)$, 
is simply the density of the system.
As is common in the analysis of bosonic Josephson junctions,
the ``survival probability'' of the system in, e.g., the left well,
is obtained by integrating the density over the left well,
$p_L(t)\equiv \frac{1}{N}\int_{-\infty}^0\rho(x,t) dx$.
Furthermore, the eigenvalues $n^{(1)}_i$ of $\rho^{(1)}(x|x';t)$ 
determine the extent to which the system is 
condensed or fragmented \cite{MCHB,ueda}. 
Finally, the first-order correlation function 
$g^{(1)}(x',x;t)\equiv\rho^{(1)}(x|x';t)/\sqrt{\rho(x,t)\rho(x',t)}$
quantifies the system's degree of spatial coherence \cite{Gla99,Sak08}.

We now turn to the details of the 1D bosonic 
Josephson junction considered in this work.
It is convenient to use dimensionless units defined
by dividing the  Hamiltonian by $\frac{\hbar^2}{mL^2}$,
where $m$ is the mass of a boson, e.g., 
$^{87}${\rm Rb} and $L$ is a length scale, e.g., $L=1\mu m$. 
One unit of energy then corresponds to $116$ {\rm Hz}. 
We have a 1D realization of an experimental 
setup similar to the one in Ref. \cite{e1} in mind.
The full many-body Hamiltonian then reads 
$H=\sum_{i=1}^N h(x_i) + \sum_{i<j}W(x_i-x_j)$,
where $h(x)=-\frac{1}{2}\frac{\partial^2}{\partial x^2} + V(x)$, 
with a trapping potential $V(x)$ and an 
interparticle interaction potential 
$W(x-x')=\lambda_0\delta(x-x')$. 

The double-well potential $V(x)$ is generated by connecting 
two harmonic potentials $V_{\pm}(x)=\frac{1}{2}(x\pm 2)^2$ 
with a cubic spline in the region $|x|\le 0.5$. 
This results in a symmetric double-well 
potential with barrier height $V(0)=1.667$. 
The lowest four single-particle energy levels 
$e_1=0.473,e_2=0.518,e_3=1.352$ and $e_4=1.611$ of 
$V(x)$  are lower than the barrier. 

Left- and right-localized orbitals $\phi_{L,R}$
can be constructed from the single-particle
ground state and the first excited state of $V(x)$.
$\phi_{L}$ and $\phi_{R}$ determine the parameters
$U=\lambda_0\int{|\phi_L|^4}$, $J=-\int{\phi_L^\ast h \phi_R}$,
the Rabi oscillation period $t_{Rabi}=\pi/J$ and the often employed
interaction parameters, $\Lambda=UN/(2J)$ and
$U/J$ \cite{t1,t2}. In this work we use 
the interaction parameter $\lambda=\lambda_0(N-1)$, 
which appears naturally in the full 
many-body treatment, and quote the 
corresponding  values for $\Lambda$ and $U/J$.
Within the framework of two-mode GP theory,
a state, which is initially localized in one well, is 
predicted to remain {\em self-trapped} 
if $\Lambda>\Lambda_c=2$ \cite{t1,t2}. 
We will consider interaction strengths
below, in the vicinity of and above $\Lambda_c$.

In all our computations the system is 
prepared at $t=0$ as the many-body ground state of the potential $V_+(x)$ 
and then propagated in the potential $V(x)$. Within the BH framework
this procedure amounts to starting from the state in which 
all bosons occupy the orbital $\phi_L$.

\begin{figure}
\includegraphics[width=8.6cm, angle=0]{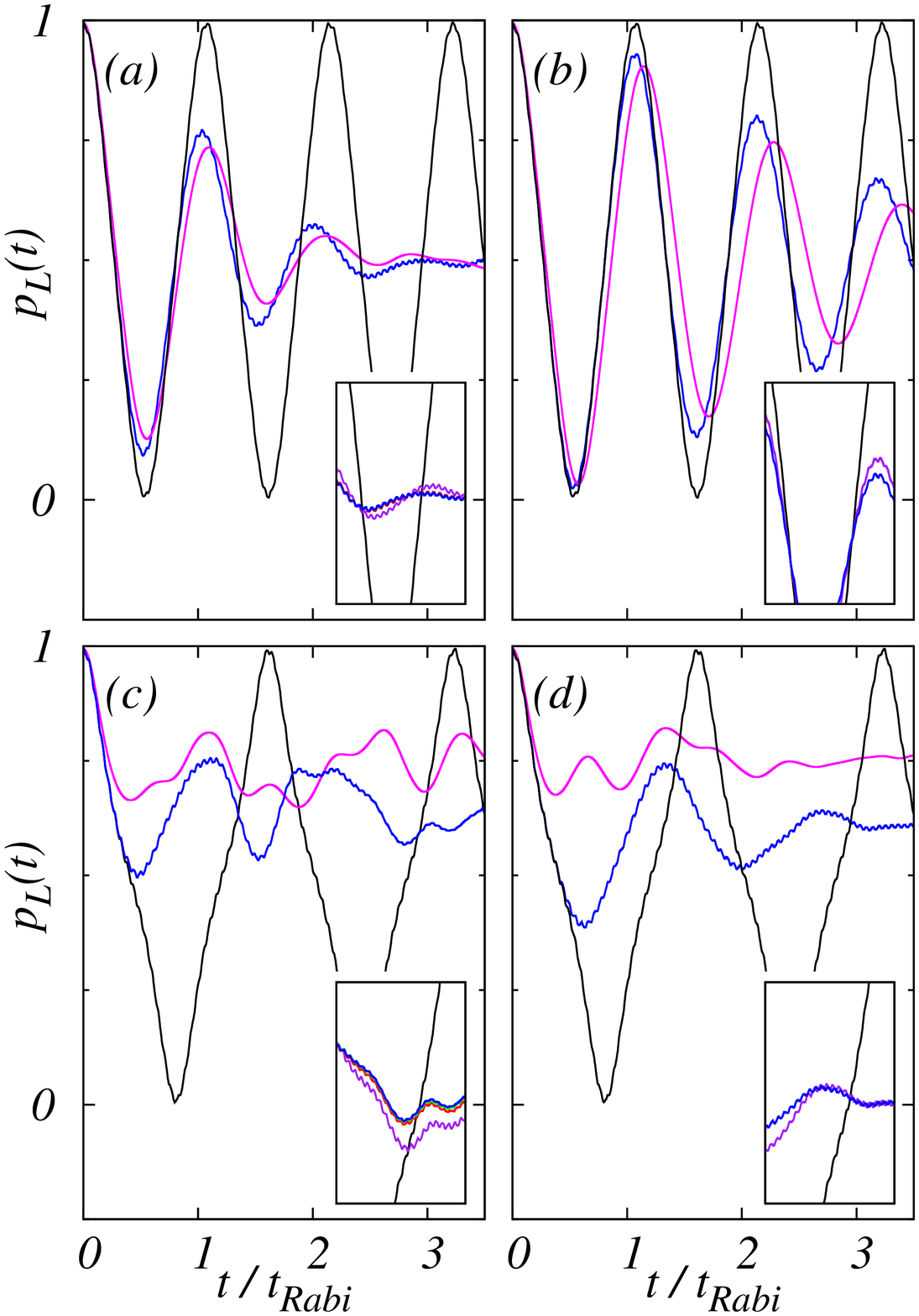}
\caption {(Color online) Full quantum dynamics of a 1D bosonic Josephson junction 
below and above the transition to self-trapping.
Shown is the full many-body result (solid blue lines) for the 
probability of finding a boson in the left well, $p_L(t)$.
For comparison, the respective GP (solid black lines) 
and BH (solid magenta lines) results are shown as well.
The parameter values are:
(a) $N=20$, $\lambda=0.152$ and (b) $N=100$, $\lambda=0.152$ (below the self-trapping transition),
(c) $N=20$, $\lambda=0.245$ and (d) $N=100$, $\lambda=0.245$ (above the self-trapping transition).
The GP and BH results are found to deviate from the full many-body results
already after short times.
The insets show the convergence of the full many-body results. (a),(c): $M=2$ (solid purple line), 
$M=4$ (solid red line), $M=6$ (solid green line), $M=8$ (solid blue line).
The $M=2$ results are seen  to
deviate slightly from the converged results for $M\ge4$.
(b),(d): The results for $M=2$ (solid purple line) and $M=4$ (solid blue line)
are shown. All quantities shown are dimensionless.
}
\label{fig1}
\end{figure}

We begin our studies with a weak interaction strength 
$\lambda=0.152$, leading to $U/J=0.140$ ($0.027$) 
and $\Lambda=1.40$ ($1.35$) for $N=20$ ($100$) bosons, 
which is well below the transition point for self-trapping $\Lambda_c=2$.
In the upper two panels of Fig.~\ref{fig1}
the full many-body (solid blue lines) results for $p_L(t)$ are
shown together with those of GP (solid black lines)
and BH (solid magenta lines) theory.
The full many-body dynamics is governed
by three different time scales. On a time scale 
of the order of a Rabi cycle,
$p_L(t)$ performs large-amplitude oscillations about $p_L=0.5$, 
the long time average of $p_L(t)$.
The amplitude of these oscillations is damped out on a 
time scale of a few Rabi cycles and marks the beginning 
of a collapse and revival (not shown) sequence \cite{t1}, 
confirmed here on the full many-body level. 
On top of these slow large-amplitude oscillations, 
a higher frequency with a small amplitude 
can be seen. In a single-particle picture these high frequency 
oscillations can be related to contributions from higher 
excited states in the initial wavefunction. 
However, a single particle picture 
fails to describe the dynamics, as we shall now show. 
While the initial wavefunction $|\Psi(t=0)\rangle$ 
is practically condensed -- the fragmentation of the 
system is less than $10^{-4}$ ($10^{-5}$) 
for $N=20$ ($100$) bosons -- the propagated wavefunction 
$|\Psi(t)\rangle$ quickly becomes fragmented. 
The fragmentation increases to about $33\%$ ($26\%$) 
at $t=3 t_{Rabi}$ for $N=20$ ($100$) particles, 
making a many-body treatment  indispensable, already 
at this weak interaction strength.
The respective GP results (solid black lines) oscillate back and forth at 
a frequency close to the Rabi frequency and resemble the full 
many-body dynamics only on a time scale {\em shorter}
than half a Rabi cycle. The poor quality of the GP mean-field 
approximation is, of course, due to the fact that the 
exact wavefunction starts to fragment while 
the GP dynamics remains condensed by construction.

The BH  (solid magenta line) 
result for $p_L(t)$ reproduces many features of that of the full many-body 
solution at this interaction strength 
for both $N=20$ and $N=100$ particles.
The large-amplitude oscillations collapse 
over a period of a few Rabi cycles 
and revive at a later stage (not shown).
Also the BH solution quickly becomes fragmented, 
starting from the left localized state, 
which is totally condensed.
The fragmentation of the BH wavefunction for $N=20$ ($100$) particles
at $t=3 t_{Rabi}$ is essentially the same 
as the respective value of the exact solution. 
However, differences between the exact and the 
BH result are visible even on time scales less than half a Rabi cycle. 
Not only are the amplitudes obviously different, but also the frequencies contained in $p_L(t)$. 
Furthermore, the BH solutions do not 
exhibit a high frequency oscillation on top of 
the slow large-amplitude oscillations; a difference 
which is related to the fact that the 
BH orbitals are {\it time-independent} and thus,
not determined variationally at each point in time. 
Note that $p_L(t)$ is a quantity in which {\em all} spatial 
degrees of freedom have been integrated out. 
Visible differences in $p_L(t)$ imply that it is not 
only the densities $\rho(x,t)$ which must differ, but also
{\em all} correlation functions.

The insets of Fig.~\ref{fig1}(a),(b) demonstrate
the convergence of the many-body dynamics results.
In particular and somewhat unexpectedly, 
the number of {\em time-dependent} orbitals
needed to describe the bosonic Josephson
junction dynamics quantitatively is $M=4$, even
below the transition point for self-trapping. 
These orbitals are determined variationally {\it at each point in time},
implying that any method using time-independent orbitals 
will need substantially more orbitals to achieve the same accuracy.

One of the central phenomena often discussed
in the context of bosonic Josephson junctions 
is the celebrated transition to self-trapping \cite{t1,t2,e1,e2}.
In what follows we would like to study the
dynamics of a bosonic Josephson junction in the self-trapping regime 
from the full many-body perspective.

The interaction strength is taken to be $\lambda=0.245$,
leading to $U/J=0.226$ ($0.043$) and $\Lambda=2.26$ 
($2.17$) for $N=20$ ($100$) particles.
Hence, the system is just above the
critical value for self-trapping $\Lambda_c=2$ \cite{t1,t2}.
The results for $N=20$ and $N=100$ are collected in Fig.~\ref{fig1}(c),(d). 
We find that the full many-body 
solutions (solid blue lines) exhibit indeed  
some self-trapping
on the time scale shown.
The  fragmentation of the condensate for $N=20$ ($100$) bosons 
increases from initially less than
$10^{-4}$ ($10^{-5}$) to about $28\%$ 
($18\%$) after three Rabi cycles.
Note that the system is now {\em less} fragmented than for weaker interactions 
after the same period of time. 
Nevertheless, GP (solid black lines) theory is -- as before -- inapplicable, 
even on time scales shorter than $t_{Rabi}/2$.
The BH (solid magenta lines) results deviate from the true 
dynamics even earlier. They greatly overestimate 
the self-trapping and coherence of the condensate. 
According to the BH model the condensate
would only be $13\%$ ($11\%$) fragmented
for $N=20$ ($100$) at $t=3 t_{Rabi}$, which is not the case.
This trend also continues for stronger interactions, see below.
The following general statement about the relationship between 
self-trapping and coherence can be inferred 
from our full many-body results:
Self-trapping is {\em only} present 
as long as the system remains coherent. 
We find this statement to be true at 
all interaction strengths and all particle numbers
considered in this work. 

We now turn to the case of stronger interactions, 
$\lambda=4.9$, which is well above the self-trapping transition point.
This leads to $U/J=9.55$ ($0.869$) and $\Lambda=47.8$ ($43.4$) 
for $N=10$ ($100$) bosons. Note that we now use 
ten instead of twenty bosons to demonstrate convergence.
The energy per particle of the full many-body 
wavefunction is now $E/N=1.22$ ($1.28$) for $N=10$ ($100$) bosons, 
which is still below the barrier height $V(0)=1.667$. 
What do we expect to happen? According to two-mode GP theory, 
the density should remain trapped in the initial well 
for any interaction strength $\Lambda\gg\Lambda_c$ \cite{t1,t2}.
Similarly, the BH model predicts ever increasing tunneling times
since the left- and right-localized states become 
eigenstates of the BH Hamiltonian in the 
limit $U/J\rightarrow\infty$ \cite{t1}.
These predictions are incorrect for stronger interactions 
since the repulsive interaction leads 
to fragmentation and broadening of the initial wavefunction,
thereby facilitating tunneling.  
The impact of the potential barrier therefore
decreases with increasing interaction strength.
We will now show that a very intricate dynamics results.

\begin{figure}[ht]
\includegraphics[width=8.6cm,angle=-0]{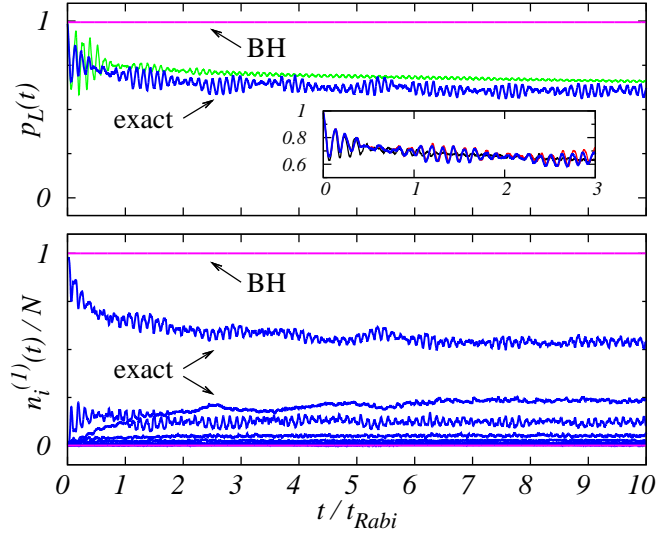}
\caption [kdv]{(Color online) 
Emergence of equilibration of the density at interaction strength
$\lambda=4.9$. Top: same as Fig.~\ref{fig1}, but for 
$N=10$ (solid blue line) and $N=100$ (solid green line). 
The respective BH (solid magenta lines) 
results are on top of each other. In contrast 
to the BH dynamics which is completely self-trapped, 
the full many-body dynamics is not.
$p_L(t)$ tends  towards its long-time average $p_L=0.5$.
For $N=100$ particles $M=4$ orbitals were used.
The inset shows the convergence of the 
full many-body solution for $N=10$ bosons: 
$M=4$ (solid black line), $M=10$ (solid blue line), $M=12$ (solid red line).
The $M=4$ result follows the trend of the converged $M=12$ result.
Bottom: corresponding natural orbital occupations
for $N=10$ bosons. The system becomes fragmented and roughly
four natural orbitals are 
macroscopically occupied.  
All quantities are dimensionless.
}
\label{fig2}
\end{figure}

Fig.~\ref{fig2}(top) shows 
the full many-body results for $N=10$ (solid blue line) 
and $N=100$ (solid green line) bosons together
with those of the BH (solid magenta line) model. 
The two BH results lie on top of each other.
In complete contrast to the BH dynamics, for which $p_L(t)$ 
remains trapped in the left well, the full many-body dynamics 
shows no self-trapping. Instead, an
equilibration phenomenon emerges,
in which the density of the system tends
to be equally distributed over both wells.

The system's full many-body dynamics is again strongly fragmented
as can be seen in Fig.~\ref{fig2}(bottom), 
which depicts the natural-orbital occupations $n_i^{(1)}$ (solid blue lines)
for $N=10$ particles. This rules out any description 
of the system by GP mean-field theory. Also shown 
are the natural-orbital occupations
of the BH (solid magenta lines) model, which  wrongly
describes a fully condensed system.

\begin{figure}[ht]
\includegraphics[width=8.6cm,angle=-0]{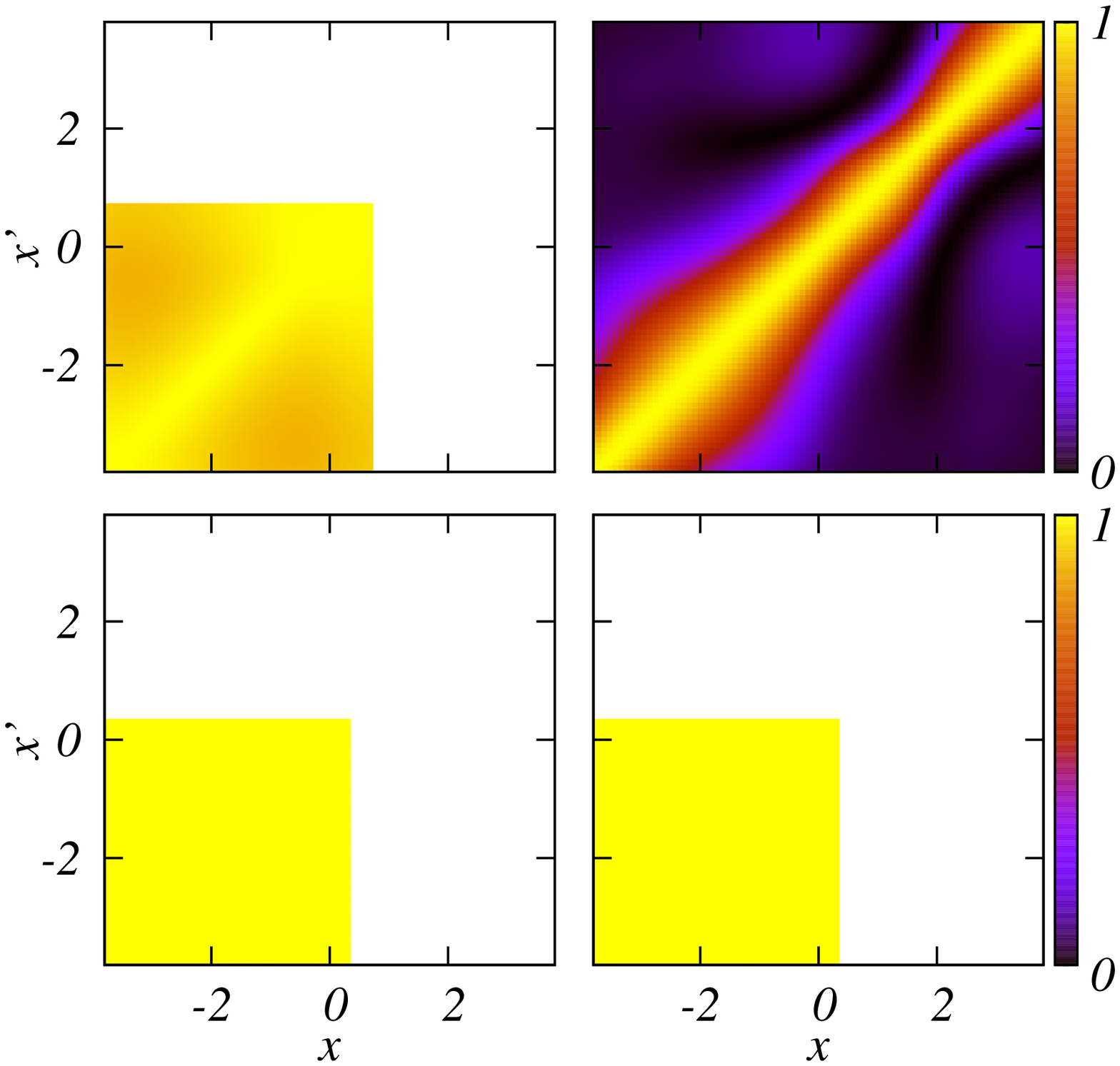}
\caption [kdv] {(Color online)
Dynamics of the first order 
correlation function for 
$\lambda=4.9$ at which the equilibration phenomenon of Fig.~\ref{fig2} occurs. 
Shown is $|g^{(1)}(x',x;t)|^2$ of $N=10$ bosons at different times. 
Top left: full many-body result at $t=0$. The initial state exhibits coherence 
over the entire extent of the system. Top right: 
full many-body result at $t=10 t_{Rabi}$. The coherence is lost. 
The system is incoherent even 
on short length scales.
Bottom left: BH result at $t=0$. Bottom right: BH result 
at $t=10 t_{Rabi}$. In contrast to the full many-body result, 
the BH wavefunction remains completely coherent. 
}
\label{fig3}
\end{figure}

The strong fragmentation of the system implies  
the presence of strong correlations. 
This can be seen in the two upper panels of Fig.~\ref{fig3}, 
which show the full many-body result for 
the first-order correlation 
function $g^{(1)}(x',x;t)$ of $N=10$ bosons at 
times $t=0$ (top left) and $t=10 t_{Rabi}$ (top right).
The fragmentation of the initial 
state is only $\approx 2\%$,
leading to an almost flat $g^{(1)}(x',x;0)$. 
This reflects the fact that the system is initially 
coherent over its entire extent. 
At $t=10 t_{Rabi}$ the coherence
of the system is completely lost
even on length scales much shorter than its size,
see upper right panel of Fig.~\ref{fig3}.
Note that also  $g^{(1)}(x',x;t)$ tends to equilibrate.
The respective BH results for $g^{(1)}(x',x;t)$ 
are shown in the two lower panels of Fig.~\ref{fig3} 
and in contrast display no visible loss of coherence.
 
Let us briefly summarize. 
We have obtained exact results for the full many-body dynamics of 
1D bosonic Josephson junctions.
The dynamics is found to be much richer 
than previously reported. In particular, 
the predictions of the commonly-employed Gross-Pitaevskii
and Bose-Hubbard theories are found to differ 
substantially from the exact results, 
already after short times and relatively weak interactions.
These differences are associated 
with the development of fragmentation and correlations
not captured by the standard theories. 
For stronger interactions, where the standard theories
predict coherence and self-trapping, we find a completely different dynamics. 
The system becomes fragmented, spatial coherence is lost 
and a long-time equilibration of the junction emerges.
We hope our results stimulate experiments.

Financial support by the DFG is acknowledged.

\begin{thebibliography}{99}
\bibitem{e4} M. R. Andrews {\it et al.}, 
 Science {\bf 275}, 637 (1997).

\bibitem{e8} T. Schumm {\it et al.},
 Nature Physics {\bf 1}, 57 (2005).

\bibitem{e35} G.-B. Jo {\it et al.},
 Phys. Rev. Lett. {\bf 98}, 030407 (2007).

\bibitem{e9} J. Est\`eve {\it et al.},
 Nature {\bf 455}, 1216 (2008).

\bibitem{e10} B. V. Hall {\it et al.},
 Phys. Rev. Lett. {\bf 98}, 030402 (2007).

\bibitem{t1} G. J. Milburn {\it et al.},
 Phys. Rev. A {\bf 55}, 4318 (1997).

\bibitem{t2} A. Smerzi {\it et al.},
 Phys. Rev. Lett. {\bf 79}, 4950 (1997).

\bibitem{e1} M. Albiez {\it et al.},
 Phys. Rev. Lett. {\bf 95}, 010402 (2005).  

\bibitem{e2} S. Levy {\it et al.}, 
 Nature, {\bf 449}, 579 (2007).

\bibitem{2mode1} S. Raghavan {\it et al.},
 Phys. Rev. A {\bf 59}, 620 (1999).

\bibitem{clee} C. Lee, Phys. Rev. Lett. {\bf 97}, 150402 (2006).

\bibitem{2mode2} D. Ananikian and T. Bergeman,
 Phys. Rev. A {\bf 73}, 013604 (2006).

\bibitem{rgat} R. Gati and M. K. Oberthaler, J. Phys. B {\bf 40}, R61 (2007).

\bibitem{meso} G. Ferrini {\it et al.},
 Phys. Rev. A {\bf 78}, 023606 (2008).

\bibitem{2mode3} X. Y. Jia {\it et al.},
 Phys. Rev. A {\bf 78}, 023613 (2008). 

\bibitem{MCTDHB} A. I. Streltsov {\it et al.}, Phys. Rev. Lett. {\bf 99}, 030402 (2007);
                 O. E. Alon {\it et al.}, Phys. Rev. A {\bf 77}, 033613 (2008).

\bibitem{parll} A. I. Streltsov {\it et al.} (to be submitted).

\bibitem{MCHB} A. I. Streltsov {\it et al.}, 
 Phys. Rev. A {\bf 73}, 063626 (2006). 

\bibitem{ueda} E. J. Mueller {\it et al.},
 Phys. Rev. A {\bf 74}, 033612 (2006).

\bibitem{Gla99} M. Naraschewski and R. J. Glauber,
 Phys. Rev. A {\bf 59}, 4595 (1999).

\bibitem{Sak08} K. Sakmann {\em et al.}, 
 Phys. Rev. A {\bf 78}, 023615 (2008). 

\end{thebibliography}
\end{document}